\documentclass[aps,prb,10pt,reprint,twocolumn,final,amsfonts,amssymb,amsmath,hypertext,superscriptaddress]{revtex4-2}
\pdfoutput=1

\usepackage{xcolor,graphicx,hyperref}
\usepackage[inline]{enumitem}
\usepackage[version=3]{mhchem}
\usepackage{bm}
\usepackage{dcolumn}
\usepackage{physics}
\usepackage{siunitx}
\usepackage{subfigure}

\ifdefined\isDraft
  
  \newcommand{\added}[1]{\textcolor{red!80!black}{#1}}
  \newcommand{\addedk}[1]{\textcolor{red!80!black}{#1}}
  \newenvironment{doAdded}{\color{red!80!black}}{\color{black}}
  \newenvironment{doAddedk}{\color{red!80!black}}{\color{black}}
\else
  
  \let\added\relax
  \let\addedk\relax
  \newenvironment{doAdded}{}{}
  \newenvironment{doAddedk}{}{}
\fi

\DeclareMathOperator{\diag}{diag}
\newcommand*{\vk}{\vb*{k}}

\graphicspath{{plot}{figure}}

\begin{document}

\preprint{APS/123-QED}

\title{Symmetry requirements for current-induced spin magnetization specific to chiral crystals: Multipole analysis and the hidden spin glide symmetry}

\author{Ryosuke Hirakida}
\affiliation{Department of Physics, the University of Tokyo, Bunkyo, Tokyo 113-0033, Japan}

\author{Masaki Kato}
\affiliation{Department of Physics, the University of Tokyo, Bunkyo, Tokyo 113-0033, Japan}

\author{Masao Ogata}
\affiliation{Department of Physics, the University of Tokyo, Bunkyo, Tokyo 113-0033, Japan}
\affiliation{Trans-Scale Quantum Science Institute, the University of Tokyo, Bunkyo, Tokyo 113-0033, Japan}

\date{\today}

\begin{abstract}
  Current-induced spin magnetization (CISM) specific to chiral crystals is microscopically analyzed using multipole theory to identify the necessary hopping and spin-orbit couplings (SOCs). Tight-binding models capturing the essence of chiral crystals are introduced to investigate the multipole degrees of freedom possessed by the Hamiltonian. The results reveal that chiral SOC has a multipole degree of freedom specific to chiral crystals. Subsequently, the CISM is evaluated numerically and analytically. The results show that in addition to the chiral SOC, hopping along the $z$-axis, which is irrelevant from a multipole perspective, is crucial for CISM. This hopping is required to break the combined symmetry of wavevector translation and spin flipping, which we refer to as \emph{spin glide symmetry}. This confirms that hopping irrelevant to chirality can play a crucial role in physical properties arising from chirality without contradicting the framework of multipole theory.
\end{abstract}

\maketitle

% ------------------------------------------------------------------------------
\section{Introduction}
\label{sec:introduction}
Materials with neither a mirror plane nor an inversion center are called chiral. This lack of symmetry causes a wide range of interesting physical properties. For instance, chirality-induced spin selectivity (CISS), a phenomenon in which a chiral material generates a spin-polarized current when an electric current is applied, has been discussed extensively~\cite{naaman2012,naaman2015,evers2022,bloom2024chiral}. CISS has been observed not only in organic molecules such as double-stranded DNA~\cite{ray1999asymmetric,gohler2011,xie2011} but also in inorganic crystals such as \ce{CrNb3S6}~\cite{inui2020chirality,nabei2020current}, \ce{NbSi2}, and \ce{TaSi2}~\cite{shiota2015chirality,shishido2021detection}. CISS in inorganic crystals has opened up further research subjects like nonlocal detection of spin polarization~\cite{shiota2015chirality,shishido2021detection}.

Several theoretical proposals have been made to explain CISS~\cite{evers2022,guo2012spin,pan2016spin,geyer2019chirality,geyer2020effective}. However, there is still no definitive theory that can completely account for these phenomena. According to simple linear response theory, spin current cannot be induced in systems with time-reversal symmetry \addedk{that have no internal orbital degrees of freedom}, but it appears in the nonlinear response regime~\cite{hirakida2022chirality} \addedk{or in systems with internal degrees of freedom~\cite{utsumi2020}}. In contrast, current-induced spin magnetization (CISM) is realized within the framework of linear response theory~\cite{yoda2015current}. Although it is natural to expect CISM when the model includes chiral hopping, it is crucial to include symmetry arguments to establish the necessary conditions for CISM.

Symmetry requirements for physical properties arising from chirality have often led to confusion because of the complexity of its symmetry~\cite{kishine2022definition}. For instance, natural optical activity and magneto-chiral effect occur in chiral materials. However, they also arise in materials belonging to strong gyrotropic point groups which contain chiral point groups as a subset~\cite{kishine2022definition}. In other words, chirality is a sufficient condition for natural optical activity and magneto-chiral effect, but it is not a necessary condition. This indicates that accurately capturing the symmetry associated with chirality is challenging.

That is why the concept of multipole~\cite{hayami2018microscopic,hayami2018classification,kusunose2020complete,hayami2020bottom,hayami2024unified} has attracted much attention as a way to analyze complex symmetries including chirality. Multipoles are a basis that considers spatial inversion \added{$\mathcal{P}$} and time-reversal symmetry \added{$\mathcal{T}$} to describe the angular dependence of electronic states. They are composed of four types according to the spatial inversion and time-reversal symmetry: electric (E) multipoles $Q_{lm}$ \added{which has the symmetry $(\mathcal{P},\mathcal{T})=((-1)^l,+)$}, magnetic (M) multipoles $M_{lm}$ \added{which has $(\mathcal{P},\mathcal{T})=((-1)^{l+1},-)$}, magnetic toroidal (MT) multipoles $T_{lm}$ \added{which has $(\mathcal{P},\mathcal{T})=((-1)^l,-)$}, and electric toroidal (ET) multipoles $G_{lm}$ \added{which has $(\mathcal{P},\mathcal{T})=((-1)^{l+1},+)$}, where $l$ and $m$ are the azimuthal and magnetic quantum number.
Since the multipoles reflect the symmetry of the system, the condition for a system to belong to a certain crystal point group can be described in terms of multipoles~\cite{hayami2018classification}. According to Tables XV and XVI in Ref.~\cite{hayami2018classification}, the condition that a system belongs to chiral crystal point groups is equivalent to the condition that its Hamiltonian has an ET monopole $G_0$, which is a $\mathcal{P}$-odd, $\mathcal{T}$-even scalar quantity.
While the chirality can be well described by multipoles, it remains unclear whether the presence of $G_0$ is a sufficient condition for the emergence of properties derived from chirality. In addition, there are still limited studies, such as Refs.~\cite{oiwa2022rotation,tsunetsugu2023theory}, that have explicitly introduced $G_0$ into models at a microscopic level.

In the present paper, we study tight-binding models for chiral crystals by introducing chiral hopping induced by spin-orbit couplings (SOCs). We analyze those models in terms of multipoles to identify the hopping required for the CISM. We show that the chiral SOC is a term with a multipole degree of freedom $G_0$ specific to chiral crystals. However, chiral SOC alone is not a sufficient condition for CISM. Surprisingly, hopping along the $z$-direction (discussed later), which is irrelevant from the viewpoint of multipole, is also necessary. This is due to a hidden symmetry, which we call \emph{spin glide symmetry}.

The remainder of the present paper is organized as follows. In Section \ref{sec:Necessary-condition}, we identify the multipole degrees of freedom required for the CISM specific to chiral crystals and introduce a chiral tight-binding model. In Section \ref{sec:Current-induced-spin-magnetization}, we evaluate the CISM of the introduced tight-binding model and identify the hopping and SOCs required for the CISM. In Section \ref{sec:Hidden-symmetry-spin}, we discuss the reason why the hopping and SOCs identified in Section \ref{sec:Current-induced-spin-magnetization} are necessary. Section \ref{sec:conclusion} is devoted to a conclusion of the present paper.

% ------------------------------------------------------------------------------
\section{Necessary condition for CISM specific to chiral crystals}
\label{sec:Necessary-condition}

\subsection{Multipole theory}

\begin{figure}[tbp]
  \centering%
  \includegraphics[width=.99\columnwidth,page=1]{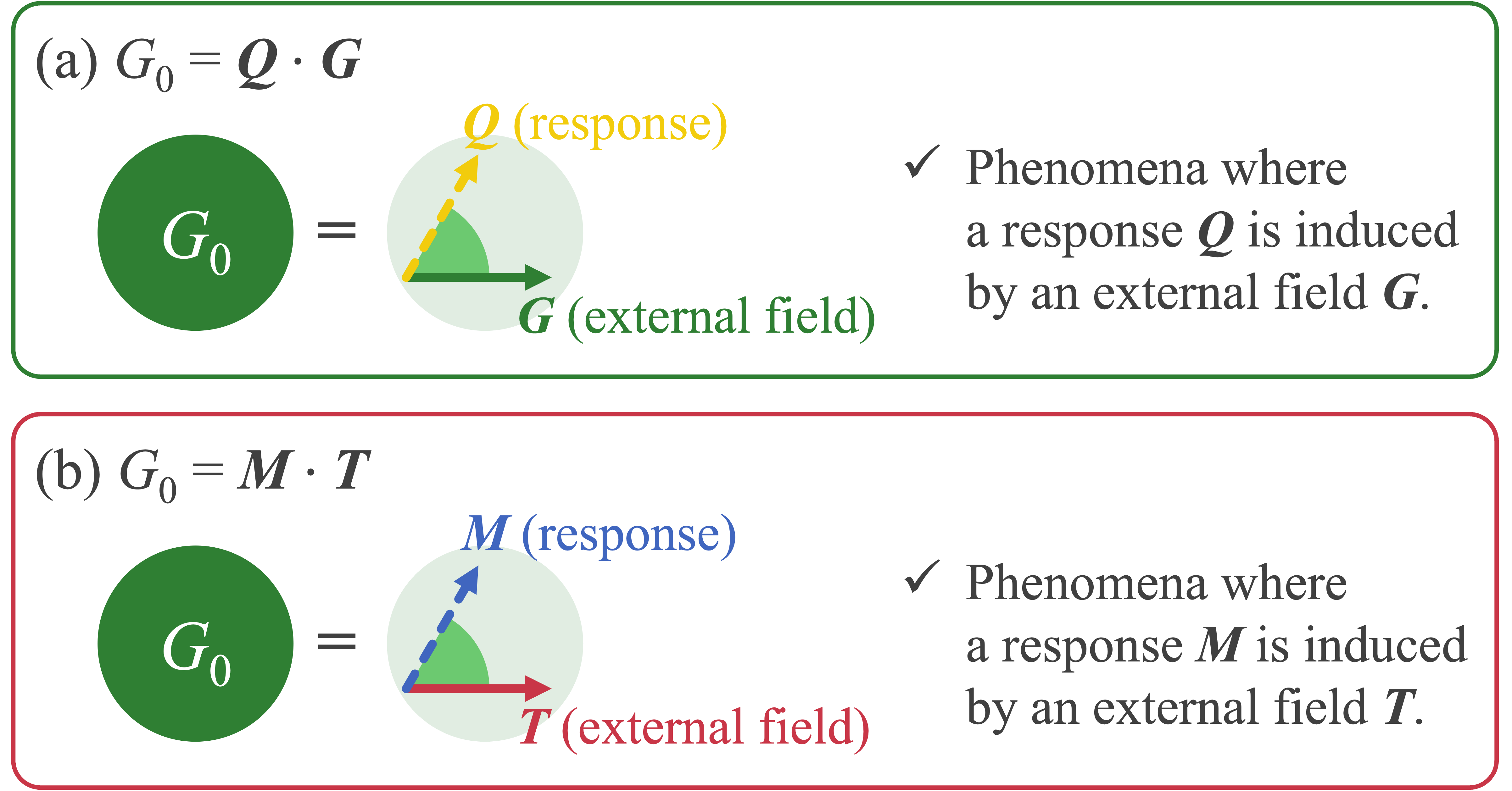}%
  \caption{%
    Two ways to express $G_0$ with dipoles and the corresponding methods to activate physical properties attributed to $G_0$. (a) The case in which $G_0$ is represented by the inner product of the ET dipole $\vb*{G}$ and the E dipole $\vb*{Q}$. For instance, this includes the phenomenon where an electric polarization $\vb*{P}$ is induced by a rotating field $\nabla\times \vb*{u}$ ($\vb*{u}$ is the lattice displacement). (b) The case in which $G_0$ is expressed by the inner product of the MT dipole $\vb*{T}$ and the M dipole $\vb*{M}$. This includes the phenomenon where a magnetization $\vb*{M}$ is induced by electric current $\vb*{j}$.
  }%
  \label{fig:how-to-activate-physical}
\end{figure}

As described in Section~\ref{sec:introduction}, multipole is an effective way to analyze complex symmetries and chirality can be characterized by the ET monopole $G_0$.
Here, we discuss that CISM is one of the possible responses originating from $G_0$. We also outline the arguments of Ref.~\cite{hayami2018classification} to confirm the necessary conditions for CISM.

Unlike other multipoles seen in classical electromagnetism, the ET multipole does not exist as a pure physical quantity but as a combination of multiple multipoles. There are two possible methods to represent the ET monopole $G_0$\added{, which has $(\mathcal{P},\mathcal{T})=(-,+)$,} using two dipoles~\cite{oiwa2022rotation}. Figure~\ref{fig:how-to-activate-physical} shows the two methods conceptually: the first is to represent $G_0$ by the inner product of the ET dipole $\vb*{G}$ \added{with $(\mathcal{P},\mathcal{T})=(+,+)$}, and the E dipole $\vb*{Q}$ \added{with $(\mathcal{P},\mathcal{T})=(-,+)$}. This means that an external field with the same symmetry as $\vb*{G}$ can induce a physical quantity with the same symmetry as $\vb*{Q}$. An example is the phenomenon of inducing an electric polarization $\vb*{P}$ (an E dipole) by a rotation field $\nabla\times \vb*{u}$ (an ET dipole), where $\vb*{u}$ is the lattice variation. This phenomenon is little-known experimentally, first proposed in Ref.~\cite{gopalan2011rotation} and microscopically investigated in Ref.~\cite{oiwa2022rotation}.

The second is to represent $G_0$ by the inner product of the MT dipole $\vb*{T}$ \added{with $(\mathcal{P},\mathcal{T})=(-,-)$}, and the M dipole $\vb*{M}$ \added{with $(\mathcal{P},\mathcal{T})=(+,-)$}. This means that an external field with the same symmetry as $\vb*{T}$ can induce a physical quantity with the same symmetry as $\vb*{M}$. An example is the phenomenon in which a magnetization $\vb*{M}$ (a M dipole) is induced by an electric current $\vb*{j}$ (a MT dipole). This phenomenon is well-known as current-induced magnetization that includes CISM and has been studied experimentally~\cite{furukawa2017observation} and theoretically~\cite{yoda2015current}. However, there have been no studies to analyze the microscopic symmetry with multipoles.

The linear response expected in the presence of a given multipole can be discussed based on the Kubo formula~\cite{hayami2018classification}.
For instance, to obtain a magnetization along the $x$- ($y$-)axis induced by a current along the $x$-axis, the Hamiltonian must have multipole degrees of freedom of either $G_0$, $G_u$ or $G_v$ ($G_{xy}$ or $Q_z$), where $G_0$ represents the ET monopole, $Q_x$, $Q_y$, and $Q_z$ represent the E dipoles, and $G_{xy}$, $G_{yz}$, $G_{zx}$, $G_u$, and $G_v$ represent the ET quadrupoles. All the necessary multipoles can be expressed symbolically by the dispersive term of the electromagnetic tensor $\hat{\alpha}^{(\mathrm{J})}$ as

\begin{align}
  \hat{\alpha}^{(\mathrm{J})}
  &=
  \begin{pmatrix}
    G_0-G_u+G_v & G_{xy}+Q_z & G_{zx}-Q_y\\
    G_{xy}-Q_z & G_0-G_u-G_v & G_{yz}+Q_x\\
    G_{zx}+Q_y & G_{yz}-Q_x & G_0+2G_u
  \end{pmatrix},
  \label{eq:multipole-electromagnetic-tensor}
\end{align}

\noindent where the matrix elements $(\hat{\alpha}^{(\mathrm{J})})_{ij}$ represent necessary multipole degrees of freedom for the magnetization along the $i$-direction to be induced when a current is along the $j$-direction ($i, j=x, y, z$).
The coefficients and signs of the multipoles in $\hat{\alpha}^{(\mathrm{J})}$ represent the anisotropy of the response. For example, consider a system with $G_v$ which is includied as $\diag(+G_v, -G_v, 0)$ in $\hat{\alpha}^{(\mathrm{J})}$.
When a current is applied in the $+x$-direction, magnetization occurs in the $+x$ (or $-x$)-direction. On the other hand, when a current is applied in the $+y$-direction, magnetization arises in the $-y$ (or $+y$)-direction with the same magnitude as in the $+x$ (or $-x$)-direction case.

Focusing on $G_0$ that characterizes the system as chiral, $G_0$ is included in the diagonal term of $\hat{\alpha}^{(\mathrm{J})}$, which means that the current-induced magnetization parallel to the applied current is one of the responses that originated from chirality. Considering this, we address in the present paper the CISM along the $z$-axis as a current-induced magnetization parallel to the applied current, which is one of the possible contributions to the spin polarization in CISS.

% ------------------------------------------------------------------------------
\subsection{Introduced tight-binding model}

\begin{figure}[tbp]
  \centering%
  \includegraphics[width=.99\columnwidth,page=1]{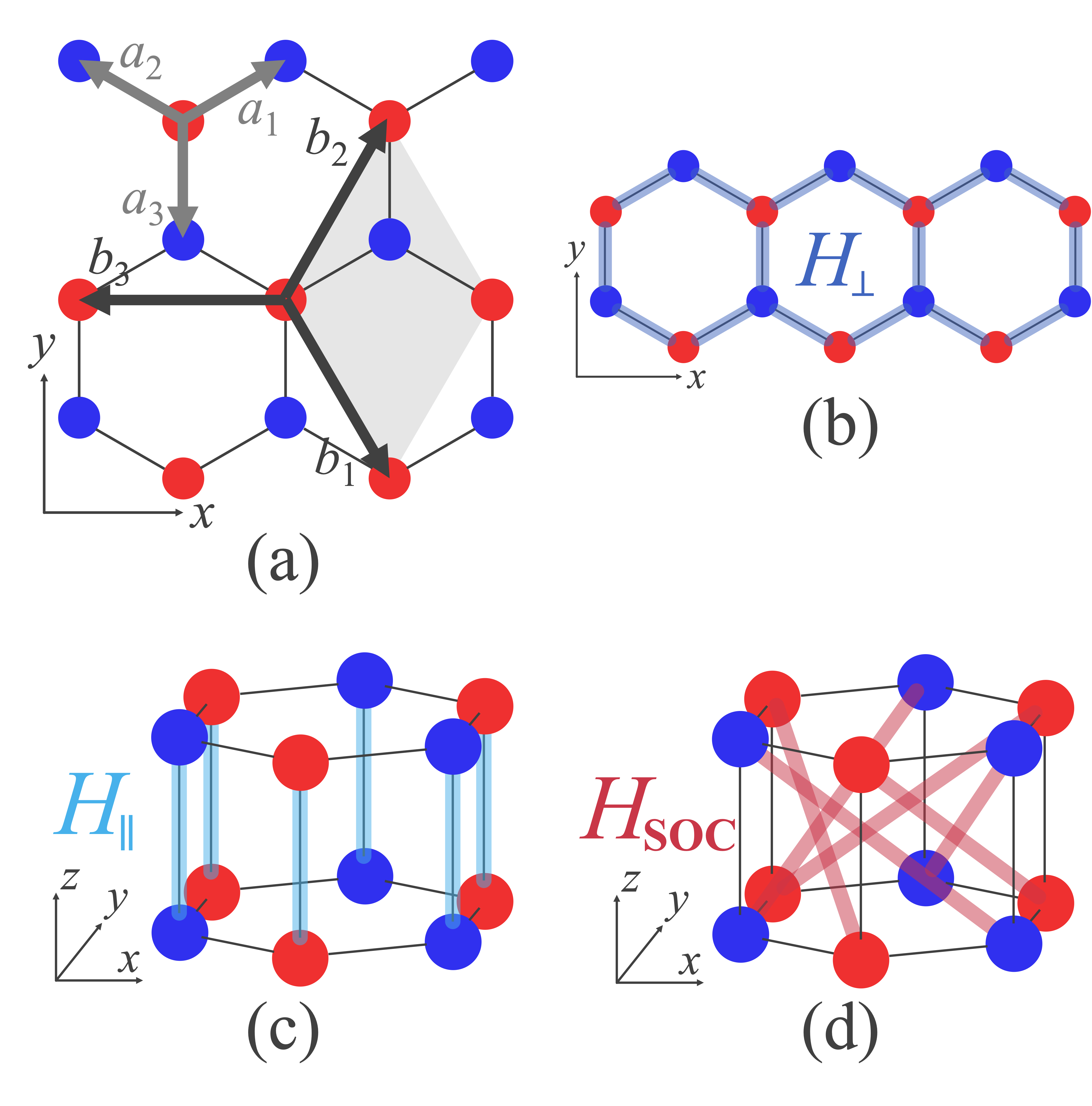}%
  \caption{%
  Tight-binding model described by the Hamiltonian in Eqs.~\eqref{eq:model-hamiltonian-first}--\eqref{eq:h_soc}.
  (a) Honeycomb layer in the $xy$-plane which is stacked along the $z$-axis. Red and blue circles represent A and B sublattices.
  (b) Nearest-neighbor hopping in the $xy$-plane.
  (c) Nearest-neighbor hopping along the $z$-axis.
  (d) Chiral hopping induced by SOCs between nearest layers. This figure is for the left-handed model ($\chi=-1$).
  }%
  \label{fig:tight-binding-model}
\end{figure}

To elucidate how the ET monopole $G_0$ leads to the CISM in chiral crystals, we present a chiral tight-binding model shown in Fig.~\ref{fig:tight-binding-model}. We show that this model is the minimal model required for the emergence of CISM.
In this model, we include a chiral hopping $H_{\mathrm{SOC}}$ induced by SOC, as shown in Fig.~\ref{fig:tight-binding-model}(d).
This hopping only allows for left-handed (or right-handed) transitions because the atomic sites, not explicitly written, are assumed to be inserted between layers.
By considering three-center integrals, we can show that $H_{\mathrm{SOC}}$ is derived from inter-atomic SOC~\cite{ozaki2023topological,kato2025interatomic}. The SOC with $G_0$ can also be derived within the framework of two-center integrals, although more complex orbital degrees of freedom are required~\cite{kato2025interatomic}.
This hopping was previously used in Ref.~\cite{yoda2015current} for evaluating spin magnetization, although the microscopic origin of spin magnetization remains unclear in Ref.~\cite{yoda2015current}.

The model is a three-dimensional tight-binding model with stacked honeycomb lattice. The Hamiltonian is written as

\begin{align}
  H
  &=\sum_{\vk}\vb*{c}^\dagger_{\vk}H(\vk)\vb*{c}_{\vk},
  \label{eq:model-hamiltonian-first}\\
  H(\vk)
  &=H_\perp(\vk)+H_\parallel(\vk)+H_{\mathrm{SOC}}(\vk),\\
  H_\perp(\vk)
  &=\sum_{a=1}^3 t_\perp\cos(\vk\cdot\vb*{a}_a)\rho_x\otimes\sigma_0
  \notag\\
  &-\sum_{a=1}^3 t_\perp\sin(\vk\cdot\vb*{a}_a)\rho_y\otimes\sigma_0,
  \label{eq:h_1}\\
  H_\parallel(\vk)
  &=2 t_\parallel\cos(\vk\cdot\vb*{c})\rho_0\otimes\sigma_0,
  \label{eq:h_3}\\
  H_{\mathrm{SOC}}(\vk)
  &=\sum_{a=1}^3 2\vb*{\lambda}_a\sin(\vk\cdot\vb*{b}_a)\cos(\vk\cdot\vb*{c})\cdot\rho_z\otimes\vb*{\sigma}
  \notag\\
  &+\sum_{a=1}^3 2\chi\vb*{\lambda}_a\cos(\vk\cdot\vb*{b}_a)\sin(\vk\cdot\vb*{c})\cdot\rho_0\otimes\vb*{\sigma}
  \label{eq:h_soc_eq_2}\\
  &+\sum_{a=1}^3 2\vb*{\mu}_a\sin(\vk\cdot\vb*{b}_a)\cos(\vk\cdot\vb*{c})\cdot\rho_0\otimes\vb*{\sigma}
  \notag\\
  &+\sum_{a=1}^3 2\chi\vb*{\mu}_a\cos(\vk\cdot\vb*{b}_a)\sin(\vk\cdot\vb*{c})\cdot\rho_z\otimes\vb*{\sigma},
  \label{eq:h_soc}
\end{align}

\noindent where $\vb*{c}_{\vk}=(c_{\vk\mathrm{X}\sigma})_{\mathrm{X}=\mathrm{A}, \mathrm{B}, \sigma=\uparrow, \downarrow}$ and $\vb*{c}^{\dagger}_{\vk}=(c^{\dagger}_{\vk\mathrm{X}\sigma})_{\mathrm{X}=\mathrm{A}, \mathrm{B}, \sigma=\uparrow, \downarrow}$ represent the annihilation and creation operators for electrons with wavenumber $\vb*{k}$, sublattice $\mathrm{X}$, and spin $\sigma$.
$\rho_0$ and $\vb*{\rho}=(\rho_\nu)_{\nu=x,y,z}$ are the $2\times2$ identity and Pauli matrices, which represent the A and B sublattice degrees of freedom.
$\sigma_0$ and $\vb*{\sigma}=(\sigma_\nu)_{\nu=x,y,z}$ are the $2\times2$ identity and Pauli matrices representing the spin degrees of freedom.
$\otimes$ is the Kronecker product of two matrices.
$H_\perp(\vk)$, $H_\parallel(\vk)$, and $H_{\mathrm{SOC}}(\vk)$ denote parts of the Hamiltonian in the wavenumber representation for the nearest neighbor hopping in the $xy$-plane, the nearest neighbor hopping along the $z$-axis, and the chiral hopping induced by SOC as shown in Fig.~\ref{fig:tight-binding-model}(b), (c), and (d), respectively.
$t_\perp$ and $t_\parallel$ are the transfer integrals for the hopping $H_\perp$ and $H_\parallel$.
$\chi=\pm1$ denotes the chirality of the model. $\chi=-1$ means that the model is left-handed, and $\chi=+1$ means that the model is right-handed.
As shown in Fig.~\ref{fig:tight-binding-model}(a), $\vb*{b}_1$ and $\vb*{b}_2$ are the lattice vectors of length $b$ for the honeycomb lattice, and $\vb*{b}_3=-(\vb*{b}_1+\vb*{b}_2)$.
$(\vb*{a}_a)_{a=1,2,3}$ are vectors of length $b/\sqrt{3}$ pointing to the nearest neighbor sites in the $xy$-plane as shown in Fig.~\ref{fig:tight-binding-model}(a).
$\vb*{c}$ is a vector of length $c$ pointing to the nearest neighbors along the $z$-axis.
$(\vb*{\lambda}_a)_{a=1,2,3}$ and $(\vb*{\mu}_a)_{a=1,2,3}$ are written as

\begin{align}
  \vb*{\lambda}_a
  &=\frac{\vb*{\lambda}_{\mathrm{A}a}+\vb*{\lambda}_{\mathrm{B}a}}{2},\\
  \vb*{\mu}_a
  &=\frac{\vb*{\lambda}_{\mathrm{A}a}-\vb*{\lambda}_{\mathrm{B}a}}{2},\\
  \vb*{\lambda}_{\mathrm{A}a}
  &=\lambda\frac{\vb*{b}_a+\vb*{c}}{\norm{\vb*{b}_a+\vb*{c}}},\\
  \vb*{\lambda}_{\mathrm{B}a}
  &=\lambda\frac{-\vb*{b}_a+\vb*{c}}{\norm{-\vb*{b}_a+\vb*{c}}}.
\end{align}

\noindent Here, $\vb*{\lambda}_{\mathrm{A}a}$ $(\vb*{\lambda}_{\mathrm{B}a})$ is the vector connecting two sites in the same A (B) sublattice on the nearest-neighbor planes as shown by red lines in Fig.~\ref{fig:tight-binding-model}(d)\addedk{, and $\lambda$ is the coupling constant for the SOC $H_{\mathrm{SOC}}$}.
In $H_{\mathrm{SOC}}$, we have assumed that the chiral hopping integrals between A (B) sublattices are proportional to $i\vb*{\lambda}_{\mathrm{A}a}\cdot\vb*{\sigma} (i\vb*{\lambda}_{\mathrm{B}a}\cdot\vb*{\sigma})$ for $\chi=-1$.
Since $\norm{\vb*{b}_a+\vb*{c}}=\norm{-\vb*{b}_a+\vb*{c}}$, we can see that $\vb*{\lambda}_a\parallel\vb*{c}$ independent of $a$, and $\vb*{\mu}_a\parallel\vb*{b}_a\ (a=1,2,3)$.

Note that $H_{\mathrm{SOC}}$ is chiral while the crystal itself is not.
Although $H_{\mathrm{SOC}}$ vanishes under the symmetry of the crystal in Fig.~\ref{fig:tight-binding-model}, it appears through a hopping considering third site via the intercalation of atomic sites between layers that are not explicitly specified, as discussed before~\cite{ozaki2023topological,kato2025interatomic}.
\added{It should also be noted that the nearest-neighbor chiral SOC hopping along the $z$-axis, $H_{\rm \parallel, SOC}(\bm{k}) = \lambda_{\parallel} \sin(\bm{k}\cdot\bm{c}) \rho_{0} \otimes \sigma_{z}$, can also be considered. However, it is impossible to introduce this hopping without significantly reducing the symmetry of the current crystal structure.}

% ------------------------------------------------------------------------------
\subsection{Multipole degrees of freedom for each hopping}
\label{sub:Multipole-degrees-of-freedom}
\begin{figure}[tbp]
  \centering%
  \includegraphics[width=.99\columnwidth,page=1]{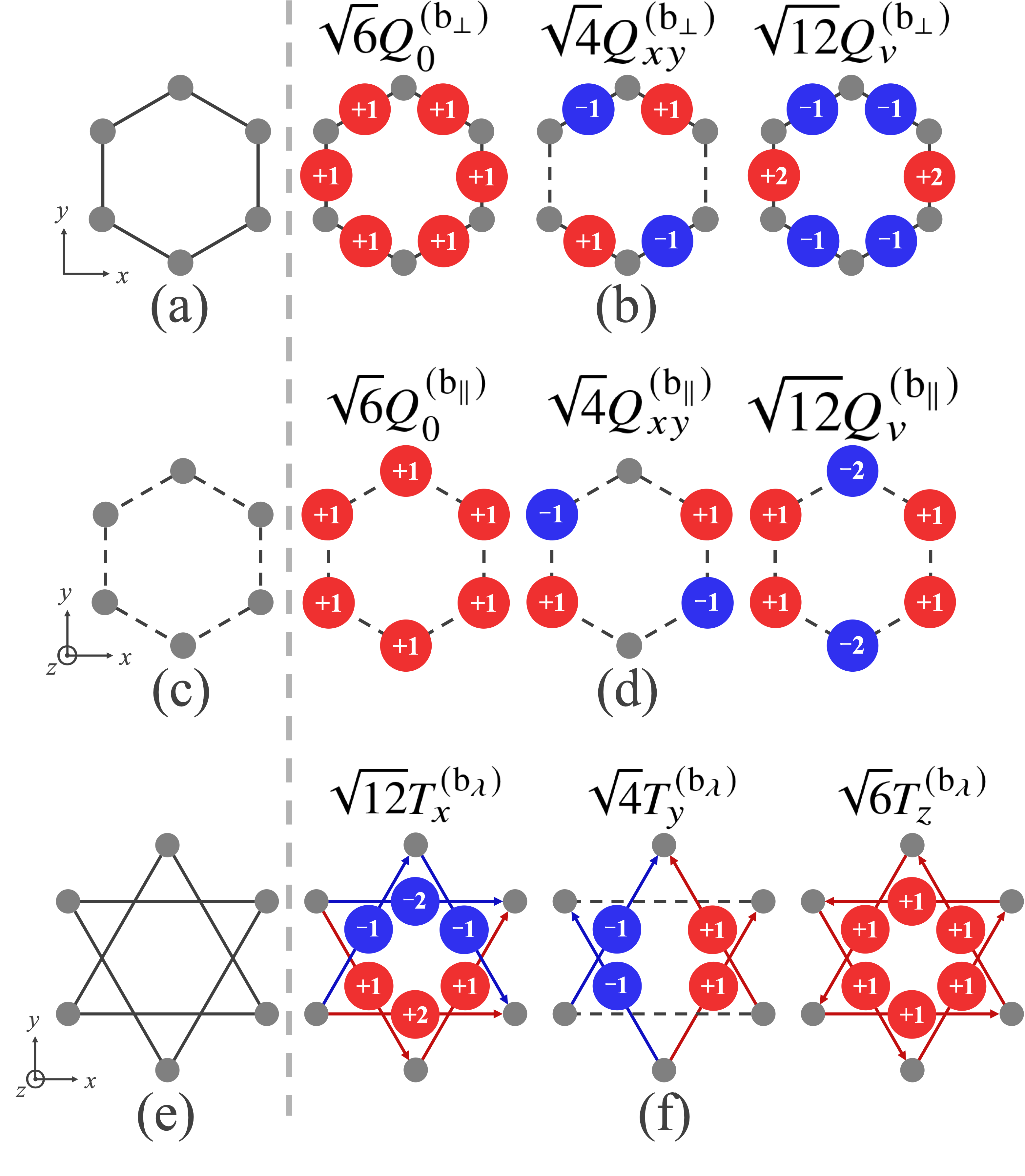}%
  \caption{%
    (a) Bond clusters composed of nearest-neighbor bonds in the $xy$-plane regenerated under the symmetric operation of $H_\perp$.
    (b) Three typical bond multipoles out of the six bond multipoles on this cluster. The red and blue circles on the bonds indicate that the real hopping is equivalent to the E monopole at the bond center. The expressions of bond multipoles on these clusters are shown above.
    \added{(c) Bond clusters composed of nearest-neighbor bonds along the $z$-axis regenerated under the symmetric operation of $H_\parallel$.}
    \added{(d) Three typical bond multipoles out of the six bond multipoles on this cluster. The labels above are the expressions of bond multipoles on these clusters.}
    (e) Bond clusters composed of next-nearest-neighbor bonds regenerated under the symmetric operation of $H_{\mathrm{SOC}}$.
    (f) Three typical bond multipoles out of the six bond multipoles on this cluster. The red and blue arrows and circles on the bonds indicate that complex hopping is equivalent to the MT dipole at the bond center.
  }%
  \label{fig:bond-multipoles}
\end{figure}

Next the multipole degrees of freedom of each hopping Hamiltonian is analyzed in the following steps by using the method of extended multipoles such as bond multipoles~\cite{hayami2018classification,hayami2020bottom,kato2025interatomic}:

\begin{enumerate}
  \item First, we examine the cluster resulting from applying all the symmetry operations of the introduced model. For $H_\perp$, $H_\parallel$ and $H_{\mathrm{SOC}}$, we use the clusters shown in Fig.~\ref{fig:bond-multipoles}(a), (c) and (e), respectively.
  Then, Fig.~\ref{fig:bond-multipoles}(b), (d) and (f) show three typical configurations out of the six basis configurations and the corresponding bond multipoles.
  Note that the coefficients of the bond multipoles have been determined for normalization~\cite{kato2025interatomic}. The MT dipoles $T_\nu\ (\nu=x,y,z)$ appear for $H_{\mathrm{SOC}}$ because the imaginary hopping on each bond, which is odd under both spatial inversion and time reversal, can be regarded as an MT dipole.
  \item Second, we express the multipoles of $H_\perp$, $H_\parallel$ and $H_{\mathrm{SOC}}$ by the tensor product of the bond multipole and spin multipole, by considering the symmetry of the hopping integrals based on the basis configurations shown in Fig.~\ref{fig:bond-multipoles}(b), (d) and (f).
  \item Finally, we expand the obtained multipoles of $H_\perp$, $H_\parallel$ and $H_{\mathrm{SOC}}$ in the basis of the composite multipoles such as $Q_0^{(\mathrm{bs})}, G_0^{(\mathrm{bs})}$ and $G_u^{(\mathrm{bs})}$~\cite{khersonskii1988quantum}.
\end{enumerate}

Detailed calculations of the multipole degrees of freedom for $H_\perp$, $H_\parallel$ and $H_{\mathrm{SOC}}$ are shown in Appendix~\ref{app:Detailed-calculations}.
Calculating the multipole degrees of freedom, we obtain

\begin{align}
  H_\perp
  &=
  \sqrt{6} Q^{(\mathrm{b_\perp s})}_0,
  \\
  H_\parallel
  &=
  \sqrt{6} Q^{(\mathrm{b_\parallel s})}_0,
  \\
  H_{\mathrm{SOC}}
  &=\added{
    \qty(\frac{\sqrt{2}c}{b}+2)G^{(\mathrm{b_\lambda s})}_0
    +\qty(\frac{2c}{b}-\sqrt{2})G^{(\mathrm{b_\lambda s})}_u
  },
\end{align}

\noindent where $Q_0$ represents the E monopole and the superscripts \added{$(\mathrm{b_\perp s})$, $(\mathrm{b_\parallel s})$ and $(\mathrm{b_\lambda s})$} denotes the composite multipoles of the bond and the spin multipoles.
Since $H_{\mathrm{SOC}}$ has $G_0$ and $G_u$, we expect a CISM parallel to the applied electric current when $H_{\mathrm{SOC}}$ is finite.

% ------------------------------------------------------------------------------
\section{Current-induced spin magnetization}
\label{sec:Current-induced-spin-magnetization}

\subsection{Numerical results}
Next, we evaluate the CISM in the model Hamiltonian in Eq.~\eqref{eq:model-hamiltonian-first}. We use the Boltzmann equation within the constant relaxation time approximation. Although the effect of interband scattering is important for the magnitude of CISM, it is not considered here because its effect does not affect the symmetry argument.

The linear response coefficient of the CISM is given by

\begin{align}
  \chi_{z;z}
  &=
  \sum_n\int_{\mathrm{BZ}}\frac{d^3k}{(2\pi)^3}
  s_{n,z}(\vk)j_{n,z}(\vk)(-\tau)f'(\varepsilon_n(\vk)),
\end{align}

\noindent where the electric field is applied in the $z$-direction.
$\int_{\mathrm{BZ}}\frac{d^3k}{(2\pi)^3}$ means the integration over the entire Brillouin zone,
% TODO: (2\pi)^3はあってる?
$\tau$ is the relaxation time,
$f(\varepsilon)$ is the Fermi distribution function, and
$\varepsilon_n(\vk)$ is the eigenenergy of the eigenstate $\ket{n\vk}$,
with band index $n$ and wavenumber $\vk$.
$s_{n,z}(\vk)=\mel{n\vk}{\hat{s}_z}{n\vk}$ is the expectation value of the spin operator $\hat{s}_z$ along the $z$-axis.
$j_{n,z}(\vk)=\frac{e}{\hbar}\pdv{\varepsilon_n(\vk)}{k_z}$ is the expectation value of the electric current operator along the $z$-axis,
where $e\ (<0)$ is the electric charge and $\hbar$ is the Dirac constant.

\begin{figure}[tbp]
  \centering%
  \includegraphics[width=.99\columnwidth,page=1]{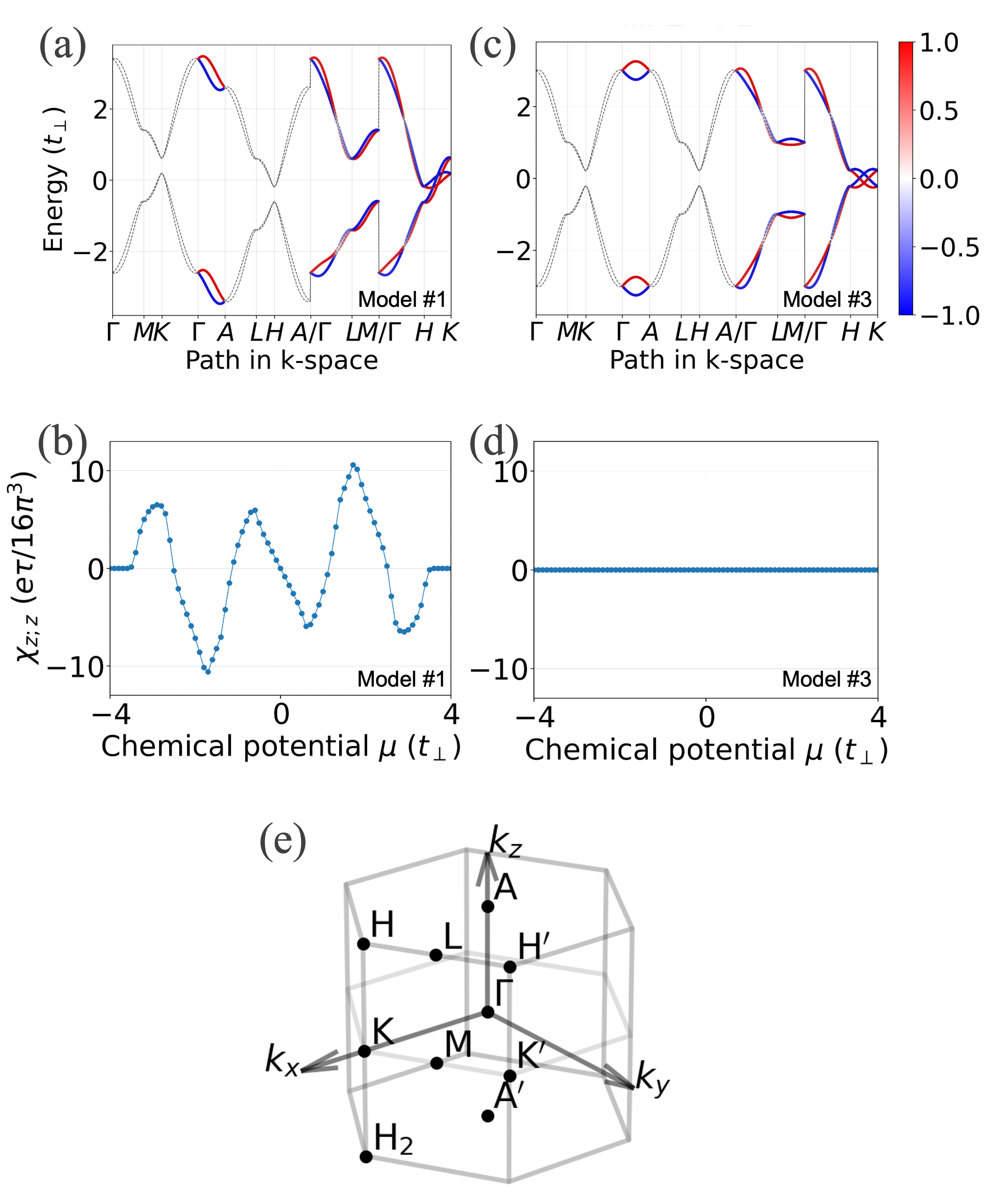}%
  \caption{%
    (a) Energy dispersion and (b) chemical potential dependence of response coefficients of the CISM for the model \#1.
    (c), (d) Those for the model \#3.
    The color bar in the energy dispersion indicates the expectation value of the spin magnetization for the corresponding states.
    Calculations are conducted for right-handed models ($\chi = +1$).
    Parameters are $t_\parallel = 0.2 t_\perp$ and $\lambda = 0.06 t_\perp$ if the corresponding term is contained in the Hamiltonian.
    (e) First Brillouin zone and representative highly symmetrical points of the introduced model.
  }%
  \label{fig:bandstructure-and-spin-magnetization}
\end{figure}

In this paper, we study the introduced model with four sets of parameters shown in Table~\ref{tab:four-model-numerical}. Figure~\ref{fig:bandstructure-and-spin-magnetization} shows the energy dispersion and the chemical potential dependence of the response coefficient of CISM in the model \#1 ($H_\perp+H_\parallel+H_{\mathrm{SOC}}$) and the model \#3 ($H_\perp+H_{\mathrm{SOC}}$). As shown in Fig.~\ref{fig:bandstructure-and-spin-magnetization}(a), the spin splitting appears when moving in the $k_z$-direction in wave number space (paths $\Gamma-\mathrm{A}, \Gamma-\mathrm{L}-\mathrm{M}$ and $\Gamma-\mathrm{H}-\mathrm{K}$) which originates from $H_{\mathrm{SOC}}$. This is also true for the energy dispersion of the model \#3 shown in Fig.~\ref{fig:bandstructure-and-spin-magnetization}(c).
Fig.~\ref{fig:bandstructure-and-spin-magnetization}(b) shows the chemical potential dependence of the finite response coefficient of the CISM of the model \#1. On the other hand, that of the model \#3 shown in Fig.~\ref{fig:bandstructure-and-spin-magnetization}(d) vanishes.

\newcommand*{\bline}{\noalign{\hrule height .08em}}
\begin{table}[tbp]
  \caption{%
  Four sets of parameters considered in the numerical calculation, and the presence or absence of CISM in each model.
  }
  \label{tab:four-model-numerical}
  \centering
  \begin{tabular}{llc}
    \bline
    No. & Hamiltonian & Response coefficient $\chi_{z;z}$ \\
    \hline
    \#1 & $H_\perp+H_\parallel+H_{\mathrm{SOC}}$ & finite \\
    \#2 & $H_\parallel+H_{\mathrm{SOC}}$ & finite \\
    \#3 & $H_\perp+H_{\mathrm{SOC}}$ & zero \\
    \#4 & $H_\perp+H_\parallel$ & zero \\
    \bline
  \end{tabular}
\end{table}

Numerical results of the presence or absence of the CISM are also shown in Table~\ref{tab:four-model-numerical}. The results in Table~\ref{tab:four-model-numerical} means that even if the Hamiltonian contains $H_{\mathrm{SOC}}$, which has chiral-specific multipole degrees of freedom $G_0$, CISM does not always appear.

% ------------------------------------------------------------------------------
\subsection{Expression of response coefficients with model parameters}
The results in Table~\ref{tab:four-model-numerical} indicate that both the hopping $H_\parallel$ and $H_{\mathrm{SOC}}$ are required for the CISM. To confirm this result analytically, we express the response coefficients of the CISM with model parameters such as the transfer integrals $t_\perp$, $t_\parallel$ and the coupling constant $\lambda$ using the method of Ref.~\cite{oiwa2022systematic}.

In Ref.~\cite{oiwa2022systematic}, a method is proposed to extract model parameters to which response coefficients are proportional.
Consider the response coefficient $\chi_{\mu;\alpha}$ of a phenomenon in which the physical quantity $\hat{A}_{\mu}$ is induced by the applied external field conjugate to the physical quantity $\hat{B}_{\alpha}$ ($\mu,\alpha=x,y,z$).
$\chi_{\mu;\alpha}$ can be expanded using the Chebyshev polynomials of the Hamiltonian, and decomposed into parts that depend on model parameters and parts that do not.
Then, one considers

\begin{align}
  \Gamma^{ij}_{\mu;\alpha}
  &=
  \sum_{\vk}\Tr[A_{\mu}(\vk)H(\vk)^iB_{\alpha}(\vk)H(\vk)^j],
\end{align}

\noindent where
$A_{\mu}(\vk)$ and $B_{\alpha}(\vk)$ are the operator of $\hat{A}_{\mu}$ and $\hat{B}_{\alpha}$ in the wavenumber representation.
If one finds the lowest order contribution of $i+j$ for which $\Gamma^{ij}_{\mu;\alpha}$ is finite, then the response coefficient $\chi_{\mu;\alpha}$ should be proportional to $\Gamma^{ij}_{\mu;\alpha}$ in the lowest order.

In the present paper, we use

\begin{align}
  \Gamma^{ij}_{z;z}
  &=
  \sum_{\vk}\Tr[s_{z}(\vk)H(\vk)^ij_{z}(\vk)H(\vk)^j],
\end{align}

\noindent where $\hat{s}_z=:\sum_{\vk}\vb*{c}^\dagger_{\vk}s_z(\vk)\vb*{c}_{\vk}$ and $j_z(\vk)=\frac{e}{\hbar}\pdv{H(\vk)}{k_z}$ are the spin and electric current operator in the wavenumber representation along the $z$-axis.
We find that the lowest order of $i+j$ is $i+j=1$, i.e.,

\begin{align}
  \Gamma^{01}_{z;z}
  &=\sum_{\vk}\Tr[s_{z}(\vk)j_{z}(\vk)H(\vk)]
  \propto t_\parallel\addedk{\chi}\lambda,
\end{align}

\noindent and $\Gamma^{10}_{z;z}=(\Gamma^{01}_{z;z})^*$. \addedk{Detailed calculations are shown in Appendix~\ref{app:Response-coefficients}.} Therefore we find that CISM is proportional to $t_\parallel\addedk{\chi}\lambda$, indicating that both $H_\parallel$ and $H_{\mathrm{SOC}}$ are necessary for a finite CISM \addedk{and that this contribution
is proportional to the chirality}.

% ------------------------------------------------------------------------------
\section{Hidden symmetry: spin glide symmetry}
\label{sec:Hidden-symmetry-spin}
In this section, we consider physical reason why not only $H_{\mathrm{SOC}}$, which has a chiral-specific multipole degree of freedom $G_0$, but also $H_\parallel$, which has nothing to do with $G_0$, is necessary for the CISM.

\begin{figure}[tbp]
  \centering%
  \includegraphics[width=.99\columnwidth,page=1]{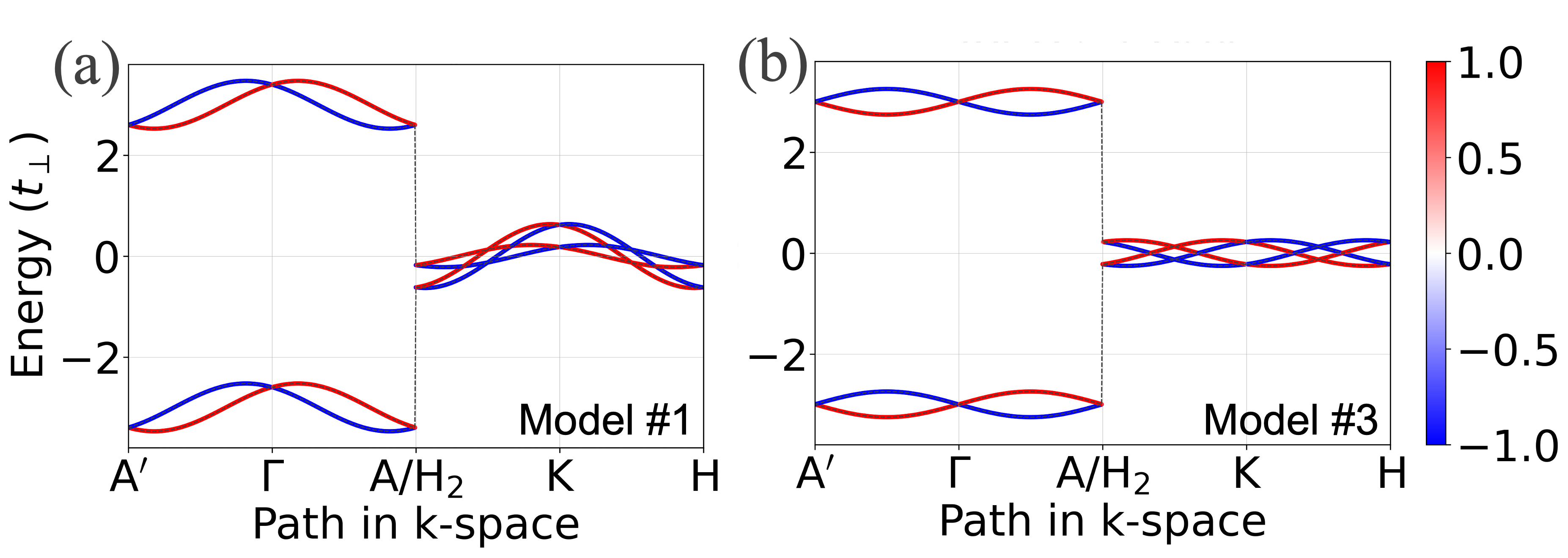}%
  \caption{%
    Energy dispersion along the $k_z$ direction (path $\mathrm{A}'-\Gamma-\mathrm{A}$ and $\mathrm{H}_2-K-\mathrm{H}$) (a) for the model \#1, and (b) for the model \#3. Calculations are conducted for right-handed models ($\chi = +1$). Parameters are $t_\parallel = 0.2 t_\perp$ and $\lambda = 0.06 t_\perp$ if the corresponding term is contained in the Hamiltonian.
  }%
  \label{fig:bandstructure-kz}
\end{figure}

The energy dispersions along the $k_z$-direction of the models \#1 and \#3 are shown in Fig.~\ref{fig:bandstructure-kz}. We can see that the energy dispersion for the model \#3 in Fig.~\ref{fig:bandstructure-kz}(b) is invariant under the combined operation of the wavenumber translation operation $k_z\to \added{k_z+\pi/c}$ and the spin-flip operation $\sigma_z\to-\sigma_z$. This can also be confirmed by the fact that the Hamiltonian $H_\perp$ and $H_{\mathrm{SOC}}$ (Eqs.~\eqref{eq:h_1} and \eqref{eq:h_soc}) are invariant under the operation $(k_z,\sigma_z)\to(\added{k_z+\pi/c},-\sigma_z)$. As a result, the summation over the whole Brillouin zone gives vanishing spin magnetization in the model \#3.

On the other hand, for the model \#1, there is no such a symmetry since the Hamiltonian $H_\parallel$ (Eq.~\eqref{eq:h_3}) is not invariant under the operation $(k_z, \sigma_z)\to(\added{k_z+\pi/c}, -\sigma_z)$. We call this notable symmetry of combined wavenumber translation and spin flipping as \emph{spin glide symmetry}. The model \#3 has spin glide symmetry, which is why the CISM vanishes even though the Hamiltonian has $G_0$.

According to the multipole theory, the CISM parallel to the applied electric current is expected when the Hamiltonian has multipole degrees of freedom $G_0$~\cite{hayami2018classification}. However, this is only a necessary condition for the CISM.
Multipoles cannot describe degrees of freedom with respect to wavenumber translation, such as spin glide symmetry.
The absence of the CISM in the model \#3 can not be predicted with the multipole theory.

% ------------------------------------------------------------------------------
\section{Conclusion}
\label{sec:conclusion}
In the present paper, we microscopically analyzed the CISM of a particular tight-binding model with multipoles to identify the hopping and SOCs required for the CISM specific to chiral crystals. The result showed that the hopping along the $z$-axis $H_\parallel$ and the chiral SOC $H_{\mathrm{SOC}}$ are necessary for the finite CISM. $H_{\mathrm{SOC}}$ is required to satisfy the necessary condition regarding the spatial inversion and time-reversal symmetries. On the other hand, $H_\parallel$ is needed to break the spin glide symmetry involving the wavenumber translation along the $z$-axis and the spin flipping which $H_{\mathrm{SOC}}$ possesses.

\added{To evaluate physical properties specific to chirality, increasing studies have been conducted that calculate the expectation value of $G_0$ in models and real materials~\cite{oiwa2022rotation,inda2024quantification}.
However, it is not sufficient to calculate the magnitude of the order parameter $G_0$ to evaluate the magnitude of the response.
A counterexample is identified, demonstrating that the presence of $G_0$ is not a sufficient condition for the physical properties specific to chirality to appear.}

% ------------------------------------------------------------------------------
\section*{Acknowledgement}
\label{sec:acknowledgement}
This work was supported by Grants-in-Aid for Scientific Research from the Japan Society for the Promotion of Science (Grant No. JP23K03274 and No. JP24KJ0580) and JST-Mirai Program (Grant No. JPMJMI19A1). R.H. was supported by World-leading Innovative Graduate Study Program for Materials Research, Information, and Technology (MERIT-WINGS).

\begin{widetext}

\appendix

\section{Detailed calculations of multipole degrees of freedom for each hopping}
\label{app:Detailed-calculations}

In this section, we calculate multipole degrees of freedom for $H_\perp$, $H_\parallel$ and $H_{\mathrm{SOC}}$ based on Refs.~\cite{hayami2018classification,hayami2020bottom,kato2025interatomic}, which can be represented as the tensor products of the bond multipoles and the spin multipoles.
\added{Here, a bond multipole is a multipole defined for hopping on the bond cluster as shown in Fig.~\ref{fig:bond-multipoles}. A spin multipole is a multipole defined in spin space. The spin multipoles of the SOCs can be represented by the corresponding Pauli matrices, while the spin multipole of the hopping without spin hybridization corresponds to the $2\times2$ identity matrix.}

In the case of $H_\perp$, three distinct bonds $\vb*{a}_i\ (i=1,2,3)$ around the most symmetric point of $H_\perp$ are shown in Fig.~\ref{fig:bond-multipoles}(a).
Because all the hopping integral in $H_\perp$ is $t_\perp$, the bond multipole is represented by $\sqrt{6}Q^{(\mathrm{b_\perp})}_0$ as shown in Fig.~\ref{fig:bond-multipoles}(b). Since the spin multipole is just $Q^{(\mathrm{s})}_0$, we obtain

\begin{align}
  H_\perp
  &=
  \sqrt{6}Q^{(\mathrm{b_\perp})}_0\otimes Q^{(\mathrm{s})}_0.
  \label{eq:h_perp_eq_q0_otimes_q0}
\end{align}

\begin{doAdded}
  \noindent where the superscripts $(\mathrm{b})=(\mathrm{b_\perp})$, $(\mathrm{b_\parallel})$ or $(\mathrm{b_\lambda})$ denote the bond multipoles and $(\mathrm{s})$ denotes the spin multipoles. We distinguish bond multipoles on different clusters and denote them as $(\mathrm{b_\perp})$, $(\mathrm{b_\parallel})$ or $(\mathrm{b_\lambda})$.

  Next, we discuss the general method for expanding multipoles in the composite multipole basis to express Eq.~\eqref{eq:h_perp_eq_q0_otimes_q0} in terms of them. The composite multipole $Z^{(\mathrm{AB})}_{l_1 l_2; lm}$ can be expressed using two distinct multipoles $Z^{(\mathrm{A})}_{l_1 m_1}$ and $Z^{(\mathrm{B})}_{l_2 m_2}$ with reference to Eq.~(74) in Ref.~\cite{kato2025interatomic} as

  \begin{align}
    Z_{l_1 l_2; lm}^{(\mathrm{AB})} = i^{l - l_1 - l_2} \sum_{m_1 = -l_1}^{l_1} \sum_{m_2 = -l_2}^{l_2} \langle l_1 m_1; l_2 m_2 | l m \rangle Z_{l_1 m_1}^{(\mathrm{A})} \otimes Z_{l_2 m_2}^{(\mathrm{B})},
    \label{eq:general_formula_composite_multipole}
  \end{align}

  \noindent where $Z$ represents either electric multipoles $Q$, magnetic multipoles $M$, magnetic toroidal multipoles $T$ or electric toroidal multipoles $G$. $Z^{(\mathrm{A})}_{l_1 m_1}$, $Z^{(\mathrm{B})}_{l_2 m_2}$, and $Z^{(\mathrm{AB})}_{l_1 l_2; lm}$ are multipoles characterized by the symmetry of the spherical harmonics $Y_{l_1 m_1}$, $Y_{l_2 m_2}$, and $Y_{lm}$, respectively. The superscripts $(\mathrm{A})$ and $(\mathrm{B})$ denote the types of the multipoles which can be $(\mathrm{b_\perp})$, $(\mathrm{b_\parallel})$, $(\mathrm{b_\lambda})$ or $(\mathrm{s})$. $\langle l_1 m_1 ; l_2 m_2 | l m \rangle$ represents the Clebsch-Gordan coefficients. $Z^{(\mathrm{AB})}_{l_1 l_2; lm}$ is determined by $l_1$, $l_2$, and $(\mathrm{AB})$, satisfying $|l_1 - l_2| \leq l \leq l_1 + l_2$.
  By combining the bond multipoles $Q^{(\mathrm{b})}_{l_1 m_1}$, $T^{(\mathrm{b})}_{l_1 m_1}$ with the spin multipoles $Q^{(\mathrm{s})}_{l_2 m_2}$, $M^{(\mathrm{s})}_{l_2 m_2}$ using Eq.~\eqref{eq:general_formula_composite_multipole} and the Clebsch-Gordan coefficients~\cite{khersonskii1988quantum}, the basis for the composite multipoles used in the following calculations can be obtained as

  \begin{align}
    Q^{(\mathrm{bs})}_{00;00}
    &=Q^{(\mathrm{b})}_{00} \otimes Q^{(\mathrm{s})}_{00},
    \label{eq:note-Q0000}
    \\
    G^{(\mathrm{bs})}_{11;00}
    &=-\qty(
      \sqrt{\frac{1}{3}}T^{(\mathrm{b})}_{11} \otimes M^{(\mathrm{s})}_{1-1}
      -\sqrt{\frac{1}{3}}T^{(\mathrm{b})}_{10} \otimes M^{(\mathrm{s})}_{10}
      +\sqrt{\frac{1}{3}}T^{(\mathrm{b})}_{1-1} \otimes M^{(\mathrm{s})}_{11}
    ),
    \label{eq:note-G1100}
    \\
    G^{(\mathrm{bs})}_{11;20}
    &=
    \sqrt{\frac{1}{6}}T^{(\mathrm{b})}_{11} \otimes M^{(\mathrm{s})}_{1-1}
    +\sqrt{\frac{2}{3}}T^{(\mathrm{b})}_{10} \otimes M^{(\mathrm{s})}_{10}
    +\sqrt{\frac{1}{6}}T^{(\mathrm{b})}_{1-1} \otimes M^{(\mathrm{s})}_{11}.
    \label{eq:note-G1120}
  \end{align}

  To use Eqs.~\eqref{eq:note-Q0000}--\eqref{eq:note-G1120}, we have to change the multipoles $\mathcal{Z}^{(\mathrm{A})}_{\nu}\ (\nu=0,x,y,z,u,\dots)$ characterized by the symmetry of the tesseral harmonics $\mathcal{Y}_{\nu}$ into the multipoles $Z^{(\mathrm{A})}_{l m}$ characterized by the symmetry of the spherical harmonics $Y_{l m}$ (where the tesseral form $\mathcal{Z}^{(\mathrm{A})}_{\nu}$ and the spherical form $Z^{(\mathrm{A})}_{l m}$ are distinguished by the presence of the calligraphic font style).
  The correspondence between $\mathcal{Z}^{(\mathrm{A})}_{\nu}$ and the spherical harmonics $Z^{(\mathrm{A})}_{l m}$ used in the following calculations can be obtained from Eqs.~(20)--(23) in Ref.~\cite{kato2025interatomic} as

  \begin{align}
    \mathcal{Z}^{(\mathrm{A})}_0&=Z^{(\mathrm{A})}_{00},\label{eq:note-Y0}\\
    \mathcal{Z}^{(\mathrm{A})}_x&=\sqrt{\frac{1}{2}}(-Z^{(\mathrm{A})}_{11}+Z^{(\mathrm{A})}_{1-1}),\\
    \mathcal{Z}^{(\mathrm{A})}_y&=i\sqrt{\frac{1}{2}}(Z^{(\mathrm{A})}_{11}+Z^{(\mathrm{A})}_{1-1}),\\
    \mathcal{Z}^{(\mathrm{A})}_z&=Z^{(\mathrm{A})}_{10},\\
    \mathcal{Z}^{(\mathrm{A})}_u&=Z^{(\mathrm{A})}_{20}.\label{eq:note-Yu}
  \end{align}

  \noindent By replacing the spherical harmonics in Eqs.~\eqref{eq:note-Q0000}--\eqref{eq:note-G1120} with the tesseral harmonics using Eqs.~\eqref{eq:note-Y0}--\eqref{eq:note-Yu}, we obtain

  \begin{align}
    Q^{(\mathrm{bs})}_0
    &=Q^{(\mathrm{bs})}_{00;00}
    =Q^{(\mathrm{b})}_0 \otimes Q^{(\mathrm{s})}_0,
    \label{eq:note-Q0}
    \\
    G^{(\mathrm{bs})}_0
    &=G^{(\mathrm{bs})}_{11;00}
    =\frac{1}{\sqrt{3}}\sum_{\nu=x,y,z}T^{(\mathrm{b})}_\nu \otimes M^{(\mathrm{s})}_\nu,
    \label{eq:note-G0}
    \\
    G^{(\mathrm{bs})}_u
    &=G^{(\mathrm{bs})}_{11;20}
    =\frac{1}{\sqrt{6}}\qty(
      3T^{(\mathrm{b})}_z \otimes M^{(\mathrm{s})}_z
      -\sum_{\nu=x,y,z}T^{(\mathrm{b})}_\nu \otimes M^{(\mathrm{s})}_\nu
    ),
    \label{eq:note-Gu}
  \end{align}

  \noindent where the original angular momenta $l_1$ and $l_2$ possessed by the composite multipole $\mathcal{Z}^{(\mathrm{AB})}_{l_1l_2;\nu}$ are omitted and simply written as $\mathcal{Z}^{(\mathrm{AB})}_{\nu}$.

  These are the general methods of expressing the composition of distinct multipoles as a basis for composite multipoles. Substituting Eq.~\eqref{eq:note-Q0} into Eq.~\eqref{eq:h_perp_eq_q0_otimes_q0}, we obtain
\end{doAdded}

\begin{align}
  H_\perp
  &=
  \sqrt{6}Q^{(\mathrm{b_\perp s})}_0.
\end{align}

In the case of $H_\parallel$, calculating the multipole in a similar way as for $H_\perp$, we obtain

\begin{align}
  H_\parallel
  &=
  \sqrt{6}Q^{(\mathrm{b_\parallel})}_0\otimes Q^{(\mathrm{s})}_0
  =
  \sqrt{6}Q^{(\mathrm{b_\parallel s})}_0,
\end{align}

\noindent \added{where the superscript $(\mathrm{b_\parallel})$ denotes the bond multipoles for the cluster of $H_\parallel$ shown in Fig.~\ref{fig:bond-multipoles}(c).}

For $H_{\mathrm{SOC}}$, we have to take care of the fact that the hopping integral in $H_{\mathrm{SOC}}$ depends on the bond direction and the spin. From Fig.~\ref{fig:tight-binding-model}(d), we can see that the hopping integral of $H_{\mathrm{SOC}}$ are

\begin{align}
  i\vb*{\lambda}_{\mathrm{A}a}\cdot\vb*{\sigma}
  &= \frac{i\lambda}{\norm{\vb*{b}_a+\vb*{c}}}(\vb*{b}_a+\vb*{c})\cdot\vb*{\sigma}
  \quad\text{(hopping of site A in the $+z$-direction)},
  \\
  i\vb*{\lambda}_{\mathrm{B}a}\cdot\vb*{\sigma}
  &= \frac{i\lambda}{\norm{-\vb*{b}_a+\vb*{c}}}(-\vb*{b}_a+\vb*{c})\cdot\vb*{\sigma}
  \quad\text{(hopping of site B in the $+z$-direction)},
  \\
  -i\vb*{\lambda}_{\mathrm{A}a}\cdot\vb*{\sigma}
  &= -\frac{i\lambda}{\norm{\vb*{b}_a+\vb*{c}}}(\vb*{b}_a+\vb*{c})\cdot\vb*{\sigma}
  \quad\text{(hopping of site A in the $-z$-direction)},
  \\
  -i\vb*{\lambda}_{\mathrm{B}a}\cdot\vb*{\sigma}
  &= -\frac{i\lambda}{\norm{-\vb*{b}_a+\vb*{c}}}(-\vb*{b}_a+\vb*{c})\cdot\vb*{\sigma}
  \quad\text{(hopping of site B in the $-z$-direction)}.
\end{align}

By using $\vb*{b}_1=\qty(\frac{b}{2}, -\frac{\sqrt{3}}{2}b, 0), \vb*{b}_2=\qty(\frac{b}{2}, \frac{\sqrt{3}}{2}b, 0), \vb*{b}_3=(-b, 0, 0)$ and $\vb*{c}=(0,0,c)$, the symmetry of $H_{\mathrm{SOC}}$ is given by

\begin{align}
  \frac{\lambda}{\norm{\vb*{b}_a+\vb*{c}}}\frac{b}{2}
  \cdot \sqrt{12}
  T^{(\mathrm{b_\lambda})}_x \otimes M^{(\mathrm{s})}_x
  \quad\text{(for the contribution of $\sigma_x$)},
  \\
  \frac{\lambda}{\norm{\vb*{b}_a+\vb*{c}}}\frac{\sqrt{3}}{2}b
  \cdot \sqrt{4}
  T^{(\mathrm{b_\lambda})}_y \otimes M^{(\mathrm{s})}_y
  \quad\text{(for the contribution of $\sigma_y$)},
  \\
  \frac{\lambda}{\norm{\vb*{b}_a+\vb*{c}}}c
  \cdot \sqrt{6}
  T^{(\mathrm{b_\lambda})}_z \otimes M^{(\mathrm{s})}_z
  \quad\text{(for the contribution of $\sigma_z$)},
\end{align}

\noindent where \added{the superscript $(\mathrm{b_\lambda})$ denotes the bond multipoles for the cluster of $H_{\mathrm{SOC}}$ shown in Fig.~\ref{fig:bond-multipoles}(e), and} the equation $\norm{-\vb*{b}_a+\vb*{c}}=\norm{\vb*{b}_a+\vb*{c}}$ is used.

Note that the spin has the symmetry of the magnetic dipole $M^{(\mathrm{s})}_\nu\ (\nu=x,y,z)$, and that the imaginary hopping integral is odd under both spatial inversion and time reversal, leading to the MT dipoles $T^{(\mathrm{b_\lambda})}_\nu\ (\nu=x,y,z)$ for bond multipoles.
Sum of these multipoles gives

\begin{align}
  H_{\mathrm{SOC}}
  &\propto
  \sqrt{3}T^{(\mathrm{b_\lambda})}_x
  \otimes
  M^{(\mathrm{s})}_x
  +\sqrt{3}T^{(\mathrm{b_\lambda})}_y
  \otimes
  M^{(\mathrm{s})}_y
  +\frac{\sqrt{6}c}{b}T^{(\mathrm{b_\lambda})}_z
  \otimes
  M^{(\mathrm{s})}_z\\
  &\added{=
  \qty(\frac{\sqrt{2}c}{b}+2)\frac{1}{\sqrt{3}}\qty(
    T^{(\mathrm{b_\lambda})}_x\otimes M^{(\mathrm{s})}_x
    +T^{(\mathrm{b_\lambda})}_y\otimes M^{(\mathrm{s})}_y
    +T^{(\mathrm{b_\lambda})}_z\otimes M^{(\mathrm{s})}_z
  )}
  \notag\\
  &\added{+\qty(\frac{2c}{b}-\sqrt{2})\frac{1}{\sqrt{6}}\qty(
    2T^{(\mathrm{b_\lambda})}_z\otimes M^{(\mathrm{s})}_z
    -T^{(\mathrm{b_\lambda})}_x\otimes M^{(\mathrm{s})}_x
    -T^{(\mathrm{b_\lambda})}_y\otimes M^{(\mathrm{s})}_y
  )}.
  \label{eq:h_soc_multipole_tesseral}
\end{align}

\noindent Here the common prefactors $\frac{\lambda b}{\norm{\vb*{b}_a+\vb*{c}}}$ is omitted.

\added{Expanding Eq.~\eqref{eq:h_soc_multipole_tesseral} in the basis of composite multipoles with Eqs.~\eqref{eq:note-G0}--\eqref{eq:note-Gu}, we can conclude}

\begin{align}
  H_{\mathrm{SOC}}
  &\added{\propto
    \qty(\frac{\sqrt{2}c}{b}+2)G^{(\mathrm{b_\lambda s})}_0
    +\qty(\frac{2c}{b}-\sqrt{2})G^{(\mathrm{b_\lambda s})}_u
  }.
\end{align}

\begin{doAddedk}
  \section{Detailed calculation of response coefficients with model parameters}
\label{app:Response-coefficients}

In this section, we calculate $\Gamma^{ij}_{z;z}=\sum_{\vk}\Tr[s_z(\vk)H(\vk)^i j_z(\vk)H(\vk)^j]$ to investigate the dependence of the response coefficient on the model parameters. To calculate $\Gamma^{ij}_{z;z}$, we calculate the following $\Omega^{ij}_{z;z}(\vk)$:

\begin{align}
  \Omega^{ij}_{z;z}(\vk)
  &=\Tr[s_z(\vk)H(\vk)^i j_z(\vk)H(\vk)^j].
  \notag
\end{align}

\noindent where $s_z(\vk)$ and $j_z(\vk)$ are expressed as

\begin{align}
  s_z(\vk)&=\frac{\hbar}{2}\rho_0\otimes\sigma_z,
  \label{eq:spin_op_wave}\\
  j_z(\vk)&=\frac{e}{\hbar}\pdv{H(\vk)}{k_z}.
  \notag
\end{align}

By using the following formulae of the Kronecker product and the Pauli matrices, we can prospectively calculate the value of $\Omega^{ij}_{z;z}(\vk)$.

\begin{align}
  (A_1\otimes B_1)(A_2\otimes B_2)
  &=(A_1A_2)\otimes(B_1B_2),
  \label{eq:a1_otimes_b1}\\
  \Tr[A_1\otimes B_1]
  &=\Tr[A_1]\Tr[B_1],\\
  %
  % \sigma_i\sigma_j
  % &=\sigma_0\delta_{ij}+i\sum_{k}\varepsilon_{ijk}\sigma_k,\\
  %
  \Tr[\sigma_i\sigma_j]
  &=2\delta_{ij},
  \label{eq:tr_sigma_i_sigma_j}
  %
  % (\vb*{a}\cdot\vb*{\sigma})(\vb*{b}\cdot\vb*{\sigma})
  % &=(\vb*{a}\cdot\vb*{b})\sigma_0+i(\vb*{a}\times\vb*{b})\cdot\vb*{\sigma}.
\end{align}

\noindent where $A_i, B_i$ ($i=1,2$) are matirces and $\delta_{ij}$ is the Kronecker delta.

In the case of $(i,j)=(0,0)$, we obtain

\newcommand*{\hsoceqii}{H^{\eqref{eq:h_soc_eq_2}}_{\mathrm{SOC}}}
\begin{align}
  \Omega^{00}_{z;z}(\vk)
  &\propto\Tr[s_z(\vk)\pdv{H(\vk)}{k_z}]
  \notag\\
  &=\Tr[s_z(\vk)\pdv{\hsoceqii(\vk)}{k_z}],
  \label{eq:tr_sz_pdv_hsoceqii}
\end{align}

\noindent where $\hsoceqii(\vk)$ is expressed using Eq.~\eqref{eq:h_soc_eq_2} as

\begin{align}
  \hsoceqii(\vk)
  &=\sum_{a=1}^3 2\chi\vb*{\lambda}_a\cos(\vk\cdot\vb*{b}_a)\sin(\vk\cdot\vb*{c})\cdot\rho_0\otimes\vb*{\sigma}.
  \notag
\end{align}

In deriving Eq.~\eqref{eq:tr_sz_pdv_hsoceqii}, it is sufficient to calculate the coefficient of the term proportional to $\rho_0\otimes\sigma_z$ in the Hamiltonian because of Eq.~\eqref{eq:spin_op_wave} and Eqs.~\eqref{eq:a1_otimes_b1}--\eqref{eq:tr_sigma_i_sigma_j}. Then,

\ifdefined\isDraft
  \newcommand{\colorRed}{\color{red}}
\else
  \let\colorRed\relax
\fi

\begin{align}
  \Omega^{00}_{z;z}(\vk)
  &\propto\sum_{a=1}^3 2\chi\lambda\cos(\vb*{k}\cdot\vb*{b}_a)
  {\colorRed\cos(\vb*{k}\cdot\vb*{c})},
  \notag\\
  \Gamma^{00}_{z;z}
  &=\sum_{\vk}\Omega^{00}_{z;z}(\vk)=0.
  \notag
\end{align}

$\Gamma^{00}_{z;z}$ vanishes because the integration of $\cos(\vb*{k}\cdot\vb*{c})$ in the Brillouin zone vanishes.

In the case $(i,j)=(0,1), (1,0)$, we obtain

% -------------------------------------------------------

\begin{align}
  \Omega^{01}_{z;z}(\vk)
  &\propto\Tr[s_z(\vk)\pdv{H(\vk)}{k_z}H(\vk)]
  \notag\\
  &=\Tr[s_z(\vk)\qty{
    \pdv{H_\parallel(\vk)}{k_z}\hsoceqii(\vk)
    +\pdv{\hsoceqii(\vk)}{k_z}H_\parallel(\vk)
  }],
  \notag\\
  \pdv{H_\parallel(\vk)}{k_z}\hsoceqii(\vk)
  &=2t_\parallel
  {\colorRed\pdv{\cos(\vb*{k}\cdot\vb*{c})}{k_z}}
  \times\sum_{a=1}^3 2\chi\vb*{\lambda}_a\cos(\vk\cdot\vb*{b}_a)
  {\colorRed\sin(\vk\cdot\vb*{c})}
  \cdot\rho_0\otimes\vb*{\sigma}
  \notag\\
  &=\sum_{a=1}^3 4 t_\parallel\chi\lambda\cos(\vk\cdot\vb*{b}_a)
  {\colorRed(-1)\sin^2(\vk\cdot\vb*{c})}
  \rho_0\otimes\sigma_z,
  \label{eq:calc_pdv_h3_hsoc}\\
  \pdv{\hsoceqii(\vk)}{k_z}H_\parallel(\vk)
  &=\sum_{a=1}^3 2\chi\vb*{\lambda}_a\cos(\vk\cdot\vb*{b}_a)
  {\colorRed\pdv{\sin(\vk\cdot\vb*{c})}{k_z}}
  \times2t_\parallel
  {\colorRed\cos(\vb*{k}\cdot\vb*{c})}
  \cdot\rho_0\otimes\vb*{\sigma}
  \notag\\
  &\propto\sum_{a=1}^3 4 t_\parallel\chi\lambda\cos(\vk\cdot\vb*{b}_a)
  {\colorRed\cos^2(\vk\cdot\vb*{c})}
  \rho_0\otimes\sigma_z,
  \label{eq:calc_pdv_hsoc_h3}\\
  \eqref{eq:calc_pdv_h3_hsoc}+\eqref{eq:calc_pdv_hsoc_h3}
  &=\sum_{a=1}^3 4 t_\parallel\chi\lambda\cos(\vk\cdot\vb*{b}_a)
  {\colorRed\cos(2\vk\cdot\vb*{c})}
  \rho_0\otimes\sigma_z,
  \notag\\
  \Gamma^{01}_{z;z}
  &=\sum_{\vk}\Omega^{01}_{z;z}(\vk)
  \propto t_\parallel\chi\lambda,
  \notag\\
  \Gamma^{10}_{z;z}
  &=(\Gamma^{01}_{z;z})^*
  \propto t_\parallel\chi\lambda.
  \notag
\end{align}

% -------------------------------------------------------

$\Gamma^{01}_{z;z}$ and $\Gamma^{10}_{z;z}$ are finite because the integral of $\cos(2\vb*{k}\cdot\vb*{c})$ within the BZ is finite. From the above, we can conclude that the lowest order of $i+j$ for which $\Gamma^{ij}_{z;z}$ is finite is in the case $i+j=1$, and in that case, $\Gamma^{ij}_{z;z}\propto t_\parallel\chi\lambda$.

\end{doAddedk}

\end{widetext}

% ------------------------------------------------------------------------------
% \bibliographystyle{apsrev4-2}
\ifdefined\isDraft
  \bibliography{lib.bib}
\else
  \bibliography{main.bbl}
\fi

\end{document}